\documentclass[journal]{IEEEtran}
\ifCLASSINFOpdf

\else

\fi
\usepackage{amsmath}
\usepackage{makeidx}  
\usepackage{algorithm}
\usepackage{algorithmic}
\usepackage{graphicx}
\usepackage{subfigure}
\usepackage{epstopdf}
\usepackage{bm}
\usepackage{cite}
\usepackage{stfloats}

\usepackage[shortlabels]{enumitem}

\usepackage{amssymb}
\usepackage{graphicx}

\usepackage{url}

\usepackage{tabularx,booktabs}
\newcolumntype{C}{>{\centering\arraybackslash}X} 
\usepackage{lipsum}

\usepackage{makecell} 

\usepackage{graphicx}
\usepackage{color}

\begin{document}

\title{Deep Learning-Based Channel Estimation for Double-RIS Aided Massive MIMO System
}

\author{\IEEEauthorblockN{Mengbing~Liu, Xin Li, Boyu Ning, Chongwen Huang, \textit{Member, IEEE},\\ Sumei Sun, \textit{Fellow, IEEE} and Chau Yuen, \textit{Fellow, IEEE}}\\
\thanks{Mengbing Liu,  Xin Li, and Chau Yuen are with Singapore University of Technology and Design, Singapore 487372 (e-mail: \tt{\{mengbingliu714, lixin.1997.lixin\}@gmail.com}; \tt{yuenchau@sutd.edu.sg})}
\thanks{Boyu Ning is with the National Key Laboratory of Science and Technology on Communications, University of Electronic Science and Technology of China, Chengdu 611731, China (e-mail: \tt{boydning@outlook.com}).}
\thanks{ Chongwen Huang is with the College of Information Science and Electronic Engineering and the Zhejiang Provincial Key Laboratory
of Information Processing, Communication and Networking, Zhejiang University, Hangzhou 310027, China, and also with the International Joint Innovation Center, Zhejiang University, Haining 314400, China (e-mail: 
\tt{chongwenhuang@zju.edu.cn}) }
\thanks{Sumei Sun is with the Institute for Infocomm Research, Agency for Science, Technology and Research, Singapore 138632 (e-mail: \tt{sunsm@i2r.a-star.edu.sg}).}
}

\maketitle
\begin{abstract}
Reconfigurable Intelligent Surface (RIS) is considered as an energy-efficient solution for future wireless communication networks due to its fast and low-cost configuration. In this letter, we consider the estimation of cascaded channels in a double-RIS aided massive multiple-input multiple-output system, which is a critical challenge due to the large number of antennas equipped at the base station and passive RIS elements. To tackle this challenge, we propose a skip-connection attention (SC-attention) network that utilizes self-attention layers and skip-connection structure to improve the channel estimation performance from the noisy pilot-based observations. Simulation results compare the proposed SC-attention network with other benchmark methods and show that SC-attention network can effectively improve the accuracy performance on normalized mean square error (NMSE) for cascaded links in a double-RIS aided system. 
\end{abstract}


\IEEEpeerreviewmaketitle

\section{Introduction}
The massive multiple-input multiple-output (MIMO) technology is one of the essential technologies in the current and next-generation wireless systems to provide high spectral efficiency\cite{bjornson2019massive}. However, the conventional massive MIMO cannot provide seamless coverage owing to some obstacles, e.g., the buildings in urban scenarios between transceivers. As a remedy, reconfigurable intelligent surface (RIS) is considered as a promising technology to recover the communication in the dead zone. Compared with other achievable methods as deploying dense micro base stations (BSs) or relays, RIS is more effective and low-cost, which can reconfigure the reflective phase
shifts via a large number of passive reflection elements without additional energy consumption \cite{huang2020holographic,wei2022multi,ning2020beamforming,huang2019reconfigurable}. 

 Due to the benefits above,  RIS has been introduced to various communication systems. To achieve better beamforming performance, RIS usually contains massive reflective elements. However, the more elements are used, the channels with larger dimension need to be estimated, which inevitably increases the overhead and complexity of channel estimation (CE) \cite{9786794,an2022scalable}. In light of this, channel state information (CSI) acquisition is one of the main bottlenecks in RIS-aided masive MIMO systems \cite{wei2021channel,ning2021terahertz,you2020wireless,zheng2021efficient}. Regarding the CE in RIS-aided systems, two iterative algorithms based on parallel factor decomposition have been introduced to efficiently reconstruct the cascaded channels \cite{wei2021channel} and a cooperative CE procedure is proposed by using beam training in terahertz RIS-adied systems \cite{ning2021terahertz}. Beyond single-reflection link, the CE problem in a cooperative double-RIS system has been analyzed in \cite{you2020wireless}. Moreover, the authors extended the work from a single-input single-output (SISO) scenario to a multi-user MIMO scenario and reformulated the CE problem to reduce the training overhead  \cite{zheng2021efficient}. The traditional CSI acquisition is mainly based on the least square (LS) algorithm with low complexity but poor performance on CSI accuracy, especially in a low signal-to-noise ratio (SNR) environment. Another classical method is  minimum mean square error (MMSE), which is widely adopted in CE problems. Besides, linear MMSE (LMMSE) has been proposed to simplify MMSE, but it still remains a performance gap with the optimal results. 

However, the cascaded channel in a RIS-aided massive MIMO system is with a high dimension due to the existence of a large number of antennas and RIS elements, and the corresponding CE complexity via conventional methods is high. To reduce the CE complexity in RIS-aided systems while remaining high CSI estimation accuracy, deep learning (DL) can be of help. Training a DL network with different channel characteristics can adapt to the changes in the environment \cite{2020,liu2021deep,9814839}. A twin conventional convolutional neural network (CNN) is designed for the estimation of the direct and cascaded channel in \cite{2020}. Based on CNN, a deep residual network has been proposed in \cite{liu2021deep} to improve the estimation performance in a RIS-assisted system.

In this letter, we propose a skip-connection attention (SC-attention) network to acquire the cascaded channels in a double-RIS aided massive MIMO network. Attention network has been used to increase the receptive field size and capacity of neural networks in image restoration \cite{wang2021uformer}. Since DL-based CE can be seen as a denoising problem which is similar to image restoration, self-attention layers can also be used for increasing the receptive field size of DL-based CSI estimation as well. The main contributions are as follows:

1) The self-attention layer has been applied to recover the CSI matrices better with a larger receptive size and model capacity. To validate its effectiveness, we visualize the denoising effect under different numbers of attention blocks.

2)By integrating skip-connection structure into our DL network, the input feature information can avoid the gradient disappearance or explosion, which improves the estimation performance significantly. Compared with the attention-only network, the accuracy of SC-attention network on normalized mean square error (NMSE) has been improved up to $16\%$. 

3) Numerical simulation results demonstrate that our proposed network outperforms both the conventional estimator (LS estimator and LMMSE estimator) and an existing DL-based network under different levels of noise, where the test data set validates the accuracy and robustness of the SC-attention network.

\textit{Notation}: Uppercase (lowercase) boldface letters denote matrix (vector) and the field of complex numbers is denoted by $\mathbb{C}$. $(\cdot)^T,(\cdot)^H $, and  $ \|\cdot\|_{\mathcal{F}}$ indicate the transpose, Hermitian, and Frobenius norm, respectively. $v \sim  \mathcal{CN}(0,\sigma^2)$ means that $v$ follows the complex Gaussian distribution with zero-mean and variance of $\sigma^2$. ${\rm{diag}}({\boldsymbol{x}})$ is a diagonal matrix whose diagonal elements are formed with elements of $\boldsymbol{x}$. $\mathbb{E}(\cdot)$ and $tr\{\cdot\}$ denote the statistical expectation and the trace, respectively.

\section{System Model}
Consider a double-RIS aided massive MIMO communication system with a BS, two RISs (referred to as RIS$_1$ and RIS$_2$), and $K$ single-antenna users. The two RISs and BS are equipped with $N$ elements and $M$ antennas. As shown in Fig. 1, we assume the direct links are blocked, two RISs are deployed near the cluster of the user and the BS, respectively. As such, users can be served by the BS via single-reflection and double-reflection links, respectively. In this paper, we assume the quasi-static flat-fading channel model for all the channels during each channel coherence interval.

Let $ \boldsymbol{h}_{kj} \triangleq \left[h_{kj,1} ,\cdots,h_{kj, N} \right]^T \in \mathbb{C}^{N \times 1}$  denote the channel between $ k $-th user and  $ {\rm{RIS}}_j$. Pilot symbols are orthogonal for distinguishing different users, i.e., $\boldsymbol{x}_p^H \boldsymbol{x}_q = 0$ for $p \neq q$. Further, $\boldsymbol{N}_{j}\in \mathbb{C}^{M \times N}$ denotes the channel from ${\rm{RIS}}_j $ to $\rm{BS}$ and $\boldsymbol{D}\triangleq \left[\boldsymbol{d}_{1} ,\cdots,\boldsymbol{d}_{N} \right] \in \mathbb{C}^{N \times N}$ denotes the channel between RISs, which only exists in the double-reflection link. Let $\boldsymbol{\theta}_j \triangleq \left[\theta_{j,1} ,\cdots,\theta_{j, N} \right]^T \in \mathbb{C}^{N \times 1}$ denote the equivalent reflection coefficients of RIS$_j$ with $ \theta_{j,n}= \beta e^{j\phi_{j,n}}$ where $\beta \in [0,1] $ and $\phi_{j,n}\in [0,2\pi)$ denotes the reflection amplitude and phase shift of the elements. The effective channel between the $k$-th user and the BS is given by 
\begin{small}
 \begin{align}
  \boldsymbol{h}_k &=  {\boldsymbol{N}}_2  \boldsymbol{\Phi}_2 \boldsymbol{D} \boldsymbol{\Phi}_1  \boldsymbol{h}_{k1} + \boldsymbol{N}_{2} \boldsymbol{\Phi}_2  \boldsymbol{h}_{k2} + \boldsymbol{N}_1 \boldsymbol{\Phi}_1  \boldsymbol{h}_{k1}\nonumber\\
   &=\sum_{n = 1}^{N} {\boldsymbol{N}}_2 {\rm{diag}}(\tilde{\boldsymbol{d}}_{k,n})\boldsymbol{\theta}_2 \theta_{1,n} + \boldsymbol{H}_{2k}\boldsymbol{\theta}_2 + \boldsymbol{H}_{1k}\boldsymbol{\theta}_1,
\end{align}
\end{small}
where $\tilde{\boldsymbol{d}}_{k,n} = \boldsymbol{d}_{n} h_{k1,n} $, and $\mathbf{\Phi}_{j} = \rm{diag}(\bm{\theta}_j)$, $\boldsymbol{H}_{1k} = \boldsymbol{N}_{1} {\rm{diag}}(\boldsymbol{h}_{k1}) \in \mathbb{C}^{M \times N} $, $\boldsymbol{H}_{2k} =\boldsymbol{N}_{2} {\rm{diag}}(\boldsymbol{h}_{k2})\in \mathbb{C}^{M \times N} $. Considering the existence of the common channel $\boldsymbol{N}_{2}$,  the expression in (1) can be re-written as
\begin{small}
 \begin{equation}
\boldsymbol{h}_k 
   =\sum_{n = 1}^{N} \boldsymbol{H}_{2k} {\rm{diag}}(\boldsymbol{h}_{3k,n})\boldsymbol{\theta}_2 \theta_{1,n} + \boldsymbol{H}_{2k}\boldsymbol{\theta}_2 + \boldsymbol{H}_{1k}\boldsymbol{\theta}_1,
\end{equation}
\end{small}
where $\boldsymbol{h}_{3k,n} = {\rm{diag}}(\boldsymbol{h}_{k2})^{-1}\tilde{\boldsymbol{d}}_{k,n} \in \mathbb{C}^{N \times 1}$ and we let $\boldsymbol{H}_{3k} = [ \boldsymbol{h}_{3k,1}, \cdots, \boldsymbol{h}_{3k,N}] \in \mathbb{C}^{N \times N}$. Then, two commonly adopted methods are introduced 
in the next subsection as benchmarks. 

 \begin{figure}[t]
\setlength{\abovecaptionskip}{0pt}
\setlength{\belowcaptionskip}{0pt}
\centering
\includegraphics[width= 0.38\textwidth]{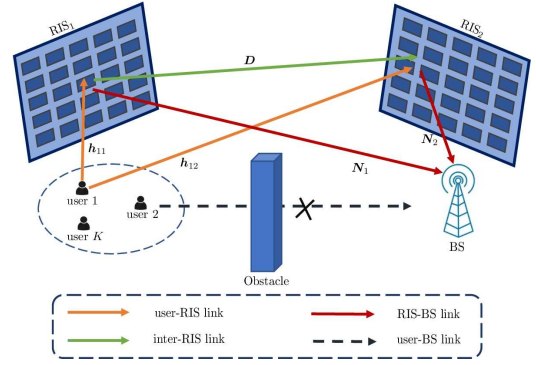}
\DeclareGraphicsExtensions.
\caption{Illustration of the Double-RIS aided system model.}
\label{1}
\vspace{-10pt}
\end{figure}

\vspace{-5pt}
\subsection{ The Estimation of Single-Reflection Links}

Considering all the elements at RIS$_2$ turned OFF\footnote{ The OFF state means that the elements are turned off and considered as part of the environment. In this state, all signals transmitted by the BS are absorbed by RISs, i.e., all the RIS elements are in a non-reflective state.}, the received signal of BS from the $k$-th user via RIS$_1$ at one time slot can be expressed as
\begin{small}
 \begin{equation}
\label{y1}
    \boldsymbol{y}_{1k} =\boldsymbol{H}_{1k} \boldsymbol{\theta}_{1}
    x_{\mathrm{1k}} +\boldsymbol{w}_{1k},
\end{equation}
\end{small}
where $x_{1k}$ and $\boldsymbol{w}_{1k} \sim  \mathcal{CN}(0,\sigma^2 \boldsymbol{I})$  denote the pilot symbol transmitted by $k$-th user and the additive white Gaussian noise (AWGN), respectively. We simply set  $x_{1k} = 1$ for ease of exposition. Thus, the received signal matrix $\boldsymbol{Y}_{1k} = [ \boldsymbol{y}_{1k}^{(1)}, \cdots, \boldsymbol{y}_{1k}^{(I_1)}]$ is given by
\begin{small}
 \begin{equation}
\label{Y1}
    \boldsymbol{Y}_{1k} =\boldsymbol{H}_{1k}  {\boldsymbol{\Phi}}_{1k} +\boldsymbol{W}_{1k},
\end{equation}
\end{small}
where ${\boldsymbol{\Phi}}_{1} = [ \boldsymbol{\theta}_{1}^{(1)},\cdots,\boldsymbol{\theta}_{1}^{(I_1)}  ]  \in \mathbb{C}^{N \times {I}_1}$ and  $\boldsymbol{W}_{1k} = [ \boldsymbol{w}_{1k}^{(1)},\cdots,\boldsymbol{w}_{1k}^{(I_1)}  ] \in \mathbb{C}^{M \times {{I}}_1}$. $I_1$ denotes the total number of pilot symbols and $I_1 \geq N$ is required to satisfy rank$\left( \boldsymbol{ \Phi}_{1}\right) = N$ for the purpose of constructing the training reflection matrix of RIS$_1$ properly. Based on (\ref{Y1}), we can obtain the CE results by the LS estimator and the LMMSE estimator.

First, the LS estimator is considered as a practical method to obtain the channel information because of its low complexity and no need for any prior knowledge of the channel \cite{zheng2021efficient}. In this case, the estimated result $\hat{\boldsymbol{H}}_{1k}^{\rm{LS}}$ is given by
\begin{small}
 \begin{equation}
   \hat{\boldsymbol{H}}_{1k}^{\rm{LS}} =\boldsymbol{Y}_{1k} {\boldsymbol{\Phi}}_{1}^H \left({\boldsymbol{\Phi}}_{1} {\boldsymbol{\Phi}}_{1}^H\right)^{-1}. 
\end{equation}
\end{small}

Next, the LMMSE estimator is developed to explore the statistical features of the channel to further improve the estimation accuracy with the help of the prior knowledge of the channel \cite{biguesh2006training}, and $\hat{\boldsymbol{H}}_{1k}^{\rm{LMMSE}}$ is given by

\begin{small}
 \begin{equation}
 \hat{\boldsymbol{H}}_{1k}^{\rm{LMMSE}} = \boldsymbol{Y}_{1k}\left({\boldsymbol{\Phi}}_{1}^H \boldsymbol{R}_{\boldsymbol{H}_{1k}}{\boldsymbol{\Phi}}_{1} + \vartheta_{1k} \boldsymbol{I}\right)^{-1} {\boldsymbol{\Phi}}_{1}\boldsymbol{R}_{\boldsymbol{H}_{1k}}, 
\end{equation}
\end{small}
where $ \boldsymbol{R}_{\boldsymbol{H}_{1k}} = \mathbb{E}(\boldsymbol{H}_{1k}^{H}\boldsymbol{H}_{1k})$ denotes the statistical channel correlation matrix of $\boldsymbol{H}_{1k}$ and $ \vartheta_{1k}= tr\{\boldsymbol{W}_{1k}^H \boldsymbol{W}_{1k}\}$. Similarly, with all the elements at $\rm{RIS}_1$ turned OFF, the estimated result $\hat{\boldsymbol{H}}_{2k}^{\rm{LS}}$ and  $\hat{\boldsymbol{H}}_{2k}^{\rm{LMMSE}}$ can be obtained.

\subsection{ The Estimation of Double-Reflection Link}

 We assume perfect cancellation of the pilot signals over the two single reflection links for ease of exposition. In this case, we consider the fixed full-ON reflection of $\rm{RIS}_2$ ($\boldsymbol{\Phi}_2 = \boldsymbol{I} $) and turn ON each element and set the phase shift as zero at $\rm{RIS}_1$ ( $\theta_{1,i} = 1$ and $\theta_{1,n} = 0, \forall n \neq i$) sequentially.  $M \geq N$ is required to satisfy the rank $(\boldsymbol{H}_{2k}) = N $ to estimate each ${\boldsymbol{h}}_{3k,n}$.  Thus, the received signal of the BS over the double-reflection link at the $n$-th time slot is given by

 \begin{small}
  \begin{align}
\boldsymbol{y}_{3k}^{(n)} &= \boldsymbol{H}_{2k}  {\rm{diag}} ({\boldsymbol{h}_{3k,n}})\boldsymbol{\theta}_2 \theta_{1, n} x_{3 k}+\boldsymbol{w}_{3 k}\nonumber\\ 
&\overset{(a)}{=} \boldsymbol{H}_{2k}  \boldsymbol{h}_{3k, n}x_{3 k} +\boldsymbol{w}_{3 k},
\end{align}
 \end{small}
 where $(a)$ follows ${\rm{diag}}({\boldsymbol{h}_{3k,n}}) \boldsymbol{\theta}_2 = \boldsymbol{\Phi}_2 \boldsymbol{h}_{3k,n}$, $x_{3k}$ and $\boldsymbol{w}_{3k} \sim  \mathcal{CN}(0,\sigma^2 \boldsymbol{I})$  denote the pilot symbol transmitted by $k$-th user and the additive white Gaussian noise, respectively.  Similarly to the last subsection, we set the pilot symbol $x_{3k} = 1 $. After $N$ time slots, the received signal matrix $\boldsymbol{Y}_{3k} = [ \boldsymbol{y}_{3k}^{(1)}, \cdots, \boldsymbol{y}_{3k}^{(N)}]$ is given by
\begin{small}
 \begin{align}
 \label{y3}
    \boldsymbol{Y}_{3k} &= \boldsymbol{H}_{2k}\boldsymbol{H}_{3k} + \boldsymbol{W}_{3k},
\end{align} 
\end{small}
 where $\boldsymbol{W}_{3k} = [ \boldsymbol{w}_{3k}^{(1)},\cdots,\boldsymbol{w}_{3k}^{(N)}  ] \in \mathbb{C}^{M \times N}$. Based on (\ref{y3}),
the estimated results $\hat{\boldsymbol{H}}_{3k}^{\rm{LS}}$ and $\hat{\boldsymbol{H}}_{3k}^{\rm{LMMSE}}$ are given by
\begin{small}
 \begin{equation}
     \hat{\boldsymbol{H}}_{3k}^{\rm{LS}} = ({\boldsymbol{H}}_{2k}^H {\boldsymbol{H}}_{2k})^{-1} {\boldsymbol{H}}_{2k}^H\boldsymbol{Y}_{3k},
 \end{equation}
\begin{equation}
  \hat{\boldsymbol{H}}_{3k}^{\rm{LMMSE}} = \boldsymbol{R}_{\boldsymbol{H}_{3k}}  ( \boldsymbol{R}_{\boldsymbol{H}_{3k}} +({\boldsymbol{H}}_{2k}{\boldsymbol{H}}_{2k}^{H} )^{-1} \vartheta_{3k} \boldsymbol{I} )^{-1}  \boldsymbol{H}_{2k}^{\dagger} \boldsymbol{Y}_{3k},
\end{equation}
\end{small}
 where $ \boldsymbol{R}_{\boldsymbol{H}_{3k}} = \mathbb{E}\left(\boldsymbol{H}_{3k}^{H}\boldsymbol{H}_{3k}\right)$ denotes the statistical channel correlation matrix of $\boldsymbol{H}_{3k}$, $ \vartheta_{3k}= tr\{\boldsymbol{W}_{3k}^H \boldsymbol{W}_{3k}\}$, and $ \boldsymbol{H}_{2k}^{\dagger} = \left( \boldsymbol{H}_{2k}^H  \boldsymbol{H}_{2k}\right)^{-1} \boldsymbol{H}_{2k}^H $. Though we can derive the estimated result based on the LS estimator and LMMSE estimator, they all have drawbacks that constrain their use in CE. As for the LS estimator, the accuracy of the result is proportional to the SNR, which means that its accuracy will be significantly affected in a low SNR environment. Though the LMMSE estimator has a better performance than the LS estimator, it still has a performance gap with the upper bound \cite{kay1993fundamentals}. Meanwhile, the intensive computation cost and the unknown end-to-end channel distribution are intractable for realizing MMSE in practice. To further improve the performance, a data-driven estimation method is developed in the next section.

\section{DL-based Channel Estimation Scheme}

Since the additive noise in the system is the main source of the estimation error, CE can be modeled as a denoising problem in this section. The SC-attention network is proposed to recover the channel coefficients from the noisy observations, i.e., the results by the LS estimator in section II. Taking advantage of the large receptive field of attention layers and the feature combination in skip-connection structure, the proposed network could further improve the estimation accuracy. 

\subsection{ Denoising Model for DL-Based Channel Estimation}

Our proposed SC-attention network regards LS results as the noisy observation, i.e., $\Tilde{\boldsymbol{Y}}_{ik}  \triangleq \hat{\boldsymbol{H}}_{ik}^{\rm{LS}
}$, and accepts them as the input at the preamble stage. Thus, the DL-based channel estimation can be modeled as a denoising problem, i.e., recovering the CSI matrix $\boldsymbol{H}_{ik}$ from a noisy observation
\begin{small}
  \begin{align}
\Tilde{\boldsymbol{Y}}_{ik} = \boldsymbol{H}_{ik} + \Tilde{\boldsymbol{W}}_{ik}, 
\label{N1}
 \end{align}
\end{small}
where $\Tilde{\boldsymbol{Y}}_{ik} = {\boldsymbol{Y}}_{ik} {\boldsymbol{\Phi}_i}^{\dagger} \in \mathbb{C}^{M \times N}, i = 1,2 $ with  ${\boldsymbol{\Phi}_i}^{\dagger} = {\boldsymbol{\Phi}}_{i}^H \left( {\boldsymbol{\Phi}}_{i} {\boldsymbol{\Phi}}_{i}^H \right)^{-1} $ and $\Tilde{\boldsymbol{Y}}_{3k} = \boldsymbol{H}_{2k}^{\dagger} \boldsymbol{Y}_{3k} \in \mathbb{C}^{N \times N} $  denote the noisy observations obtained by the LS estimator. $\Tilde{\boldsymbol{W}}_{ik} = \boldsymbol{W}_{ik} {\boldsymbol{\Phi}_i}^{\dagger}  \in \mathbb{C}^{M \times N}, i = 1,2 $ and $\Tilde{\boldsymbol{W}}_{3k} =  \boldsymbol{H}_{2k}^{\dagger} \boldsymbol{W}_{3k} \in \mathbb{C}^{N \times N}$ are the noise components. 
  \begin{figure}[t]
\setlength{\abovecaptionskip}{0pt}
\setlength{\belowcaptionskip}{0pt}
\centering
\includegraphics[width= 0.43\textwidth]{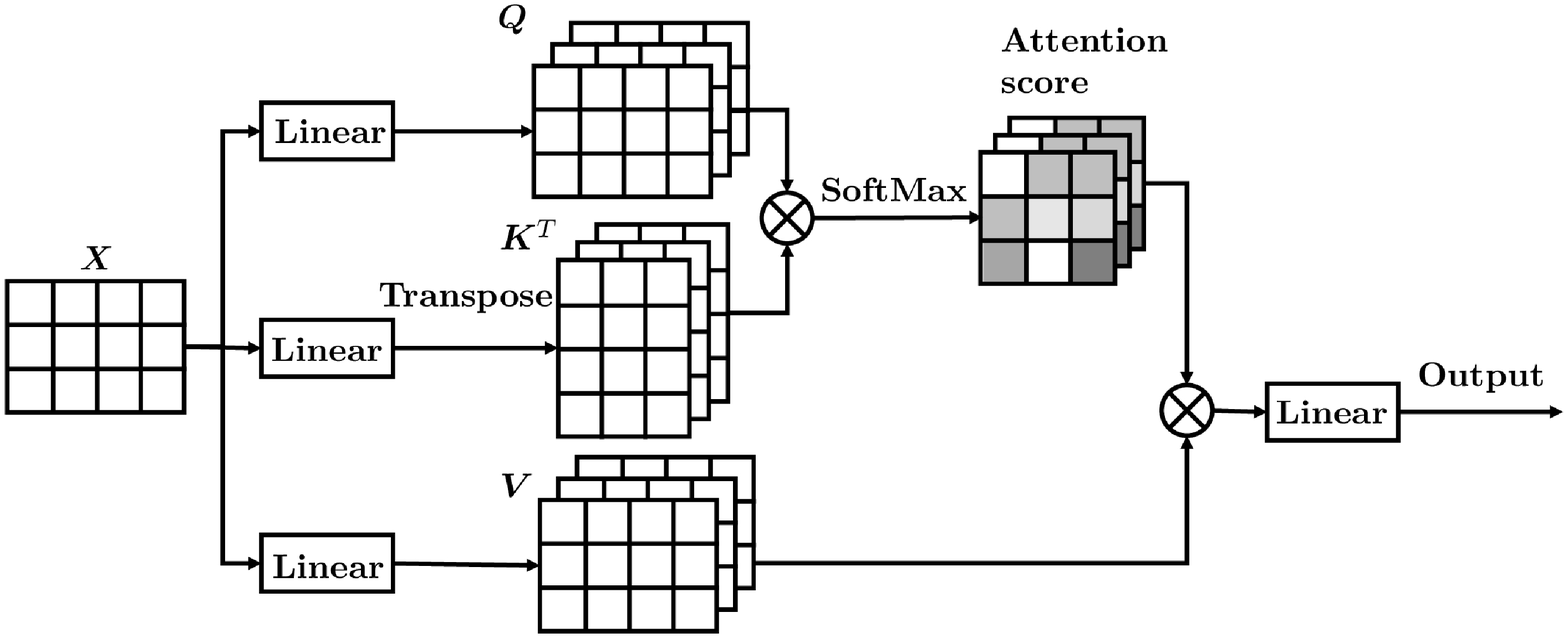}
\DeclareGraphicsExtensions.
\caption{Overview of the self-attention layer.}
\label{2}
\vspace{-10pt}
\end{figure}
Then, the input-label pair $(\Tilde{\boldsymbol{Y}}_{ik},\boldsymbol{H}_{ik}), i = 1,2,3 $ is used as one sample for the training process of the three links, respectively. Thus, the whole data set for the $i$-th link with $T$ samples can be formulated as
 \begin{small}
  \begin{align}
(\tilde{\mathcal{Y}}_i, \mathcal{H}_i)=   \left\{ (\Tilde{\boldsymbol{Y}}_{ik}^{(1)}, \boldsymbol{H}_{ik}^{(1)} ),\cdots, ( \Tilde{\boldsymbol{Y}}_{ik}^{(T)}, \boldsymbol{H}_{ik}^{(T)}) \right\}.
 \end{align}
 \end{small}

\begin{figure*}[ht]
\centering
\includegraphics[scale=.35]{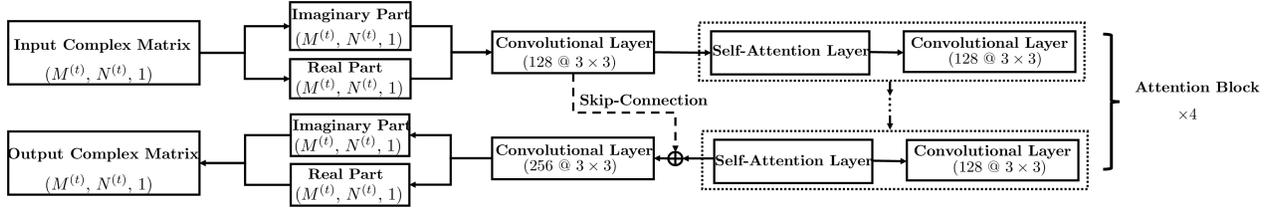}
\caption{The structure of proposed SC-attention network. }
\label{3}
\end{figure*}
\vspace{-6pt}

\subsection{Self-Attention Layer}

Self-attention layers are utilized to have a large receptive field spanning the whole input, with less computational complexity than when using large convolutional kernels  \cite{vaswani2017attention}. 
The structure of the self-attention layer is shown in Fig. \ref{2}, which is based on the scaled dot-product attention. The self-attention layer comprises four fully connected layers (FCLs) and two matrix multiplications. First, the input feature map $\boldsymbol{X}$ is passed through three separate FCLs to generate the key map $\boldsymbol{K}$, query map $\boldsymbol{Q} $, and value map $\boldsymbol{V} $, respectively. $\boldsymbol{K}$ is transposed and multiplied with $\boldsymbol{Q} $, resulting in the dot-product $\boldsymbol{A}$. After a SoftMax function, the attention map $\boldsymbol{A}^{\prime}$ with a scaled attention score can be expressed as
\begin{small}
\begin{align}
\boldsymbol{A}^{\prime} = {\rm{SoftMax}}\left(\boldsymbol{K}^T\boldsymbol{Q}\right),
\end{align}
\end{small}
where the SoftMax function is defined as $\boldsymbol{\alpha}_{ij}^{\prime} =\exp \left(\alpha_{ij}\right) / \sum_{i} \sum_{j} \exp \left(\alpha_{ij}\right)$ with the attention score between $\alpha_{ij}$ and other elements in  $\boldsymbol{A}$. Finally, the matrix multiplication of the value and attention map is sent to an FCL and the result is used as the output. Thus, self-attention layers have been used in our network to better model the CSI characteristics.

\subsection{Overall Structure of SC-Attention Network}
The overall structure of the SC-attention network is shown in Fig. \ref{3}, where the skip-connection structure has been applied to improve the network performance \cite{ronneberger2015u}. As mentioned in the previous section, self-attention layers enable neural networks to have a large receptive field to model the CSI characteristics better. Furthermore, the skip-connection structure can accelerate the training process and avoid gradient disappearance or gradient explosion by combining the input features into the output of the whole attention structure.

The structure of the SC-attention network is elaborated as follows. First, the rough CSI matrix $\Tilde{\boldsymbol{Y}}_{ik}$ of dimension $(M^{(t)},N^{(t)},1)$ is used as the input of the network where $M^{(t)}= M, N^{(t)} = N$ for single-reflection links and $M^{(t)}= N^{(t)} = N $ for the double-reflection link. In order to feed the deep network, the input complex matrix $\Tilde{\boldsymbol{Y}}_{ik}$ can be divided into real and imaginary parts with the same dimension  and then concatenated into a real-valued matrix of dimension $(M^{(t)}, N^{(t)}, 2)$ \cite{dong2019deep}. The first convolutional layer uses 128 filters of size $3 \times 3 $ followed by a ReLU activation. Then, the matrix is used as the input of four attention blocks, and each block consists of a self-attention layer and a convolutional layer with the same size as the first layer. Meanwhile, the feature information of the first convolutional  layer is concatenated to the output of the attention block by the skip-connection structure to enhance the network performance. After that, the mixed matrix is sent to a convolutional layer with 256 filters of size $3 \times 3$, and the output is then reshaped into a dimension of $(M^{(t)}, N^{(t)}, 2)$ again. Finally, the matrix is concatenated into the output complex matrix of $(M^{(t)}, N^{(t)},1)$.  

As for the training details, Adam optimizer \cite{kingma2014adam} is adopted for the training process, with a total of 100 epochs on one NVIDIA RTX 2080Ti GPU. The initial learning rate, weight decay, and batch size are set to $1 \times 10^{-3}$, $1 \times 10^{-5}$, and 64 respectively. In addition, the NMSE metric is adopted as the loss function, i.e., the loss between ${\boldsymbol{H}}_{ik}$ and estimated result by SC-attention network ${\hat{\boldsymbol{H}}}_{ik}$ is defined as $\frac{1}{T} \sum_{t=1}^{T}\left\|\boldsymbol{H}_{ik}^{(t)}-{\hat{\boldsymbol{H}}}_{ik}^{(t)}\right\|_{\mathcal{F}}^2 /\left\|\boldsymbol{H}_{ik}^{(t)}\right\|_{\mathcal{F}}^2$.

\section{Simulation Results}

In this section, numerical simulations are provided to verify the effectiveness of SC-attention network. Throughout the simulation, the number of antennas $M$ is 64, and the number of RIS elements $N$ is 32. Generally, all the channels are assumed to follow the independent and identically distributed Racian fading channel model, i.e., $ \boldsymbol{H} = \sqrt{\gamma/(\gamma+1)} \bar{\boldsymbol{H}}+\sqrt{{1}/(\gamma+1)}\Tilde{\boldsymbol{H}}$, where $\bar{\boldsymbol{H}}$, $\Tilde{\boldsymbol{H}}$ are the line-of-sight (LoS) and non-LoS (NLoS) components, and $\gamma$ is the Rician factor of the channel. In addition, the path loss can be modeled as $\beta = \beta_0 (d/d_0)^{-\alpha}$, 
where $\beta_0 = -15$ $\mathrm{dB}$ is the path loss at the reference distance $d_0 = 10m$ and $\alpha$ is the path loss exponent. Thus, the path loss factor can be taken into account by multiplying $\boldsymbol{H}$ with $\sqrt{\beta}$ and obtain the attenuation of each link. As such, the total attenuation of single-reflection links and double reflection link are $\boldsymbol{N}_{2} \boldsymbol{\Phi}_2  \boldsymbol{h}_{k2}  (\boldsymbol{N}_1 \boldsymbol{\Phi}_1  \boldsymbol{h}_{k1})$ and ${\boldsymbol{N}}_2  \boldsymbol{\Phi}_2 \boldsymbol{D} \boldsymbol{\Phi}_1  \boldsymbol{h}_{k1}$.
The simulation parameters in this paper are set as follows. The Rician factors are set as $\gamma_{\boldsymbol{h}_{k1}}=0$, $\gamma_{\boldsymbol{h}_{k2}} = 10$, $ \gamma_{\boldsymbol{D}} = 10 $, $ \gamma_{\boldsymbol{N}_2} = 0$, $ \gamma_{\boldsymbol{N}_1} = 10$. The corresponding distances are $d_{\boldsymbol{h}_{k1}} = 16$ m, $d_{\boldsymbol{h}_{k2}} = 90$ m, $d_{\boldsymbol{D}} = 80$ m, $d_{\boldsymbol{N}_2} = 16$ m, $d_{\boldsymbol{N}_1} = 90$ m. Additionally, the path loss exponents are $\alpha_{s_1} = 2.3, s_1 \in \{{\boldsymbol{h}_{k2}}, {\boldsymbol{D}},{\boldsymbol{N}_1}\} $ and $\alpha_{s_2} = 2, s_2 \in \{{\boldsymbol{h}_{k1}}, {\boldsymbol{N}_2}\} $. The SNR in the simulation is set as the transmit SNR which is defined as $\mathrm{SNR} = P/\sigma^2$. The total number of samples is $T = 120000$ where $80 \% $  and $20 \%$ of the whole generated data are used for training and validation, respectively. Both the training and the validation are performed with the SNR from -10dB to 15dB to illustrate the generality of our network. To see the performance, We use NMSE as the metric to evaluate the accuracy of the estimated results by the LS estimator, LMMSE estimator, SF-CNN\cite{2020} and SC-attention network. 
\begin{figure}[!ht]
\setlength{\abovecaptionskip}{-5pt}
\setlength{\belowcaptionskip}{0pt}
\centering
\includegraphics[width= 0.4\textwidth]{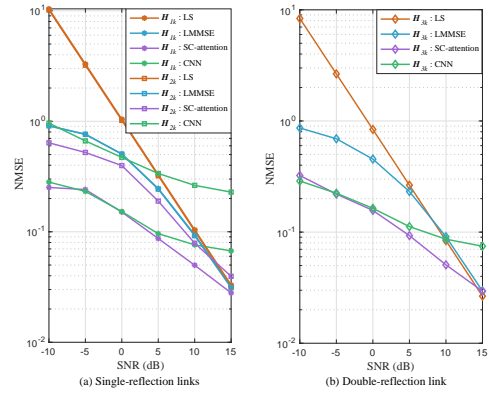}
\DeclareGraphicsExtensions.
\caption{Channel estimation NMSE with respect to SNR.}
\label{snr}
\vspace{-5pt}
\end{figure}

The channel estimation for single-reflection links $(\boldsymbol{H}_{1k}$, $\boldsymbol{H}_{2k})$ and double-reflection link $(\boldsymbol{H}_{3k})$ with respect to SNR are presented in Fig. \ref{snr}(a) and Fig. \ref{snr}(b), respectively. We can see that the LS estimator and LMMSE estimator have the same results on both $\boldsymbol{H}_{1k}$ and $\boldsymbol{H}_{2k}$ due to the same single-reflection channel characteristic in Fig. \ref{snr}(a). Our proposed SC-attention network is superior to the conventional methods, i.e., LS estimator and LMMSE estimator. This is because the LMMSE estimator is linear whereas the proposed SC-attention method is non-linear and powerful in constructing non-linear mapping by intelligently exploiting the non-linear spatial features between the input and output.
Moreover, our proposed SC-attention network performs better than the SF-CNN baseline due to their better mapping architectures from the noise observation to channel data.  We also observe that the performance of the SC-attention is similar to the other conventional estimators at high SNR (i.e., 15 dB). According to (\ref{N1}), the observation with less noise means that the gap between the input and the label is becoming narrow. Thus, it is easy to understand that our proposed SC-attention network performs even worse under high SNR condition because of the biased nature of the neural networks which do not provide unlimited accuracy. 
 \begin{figure}[!ht]
\setlength{\abovecaptionskip}{0pt}
\setlength{\belowcaptionskip}{0pt}
\centering
\includegraphics[width= 0.4\textwidth]{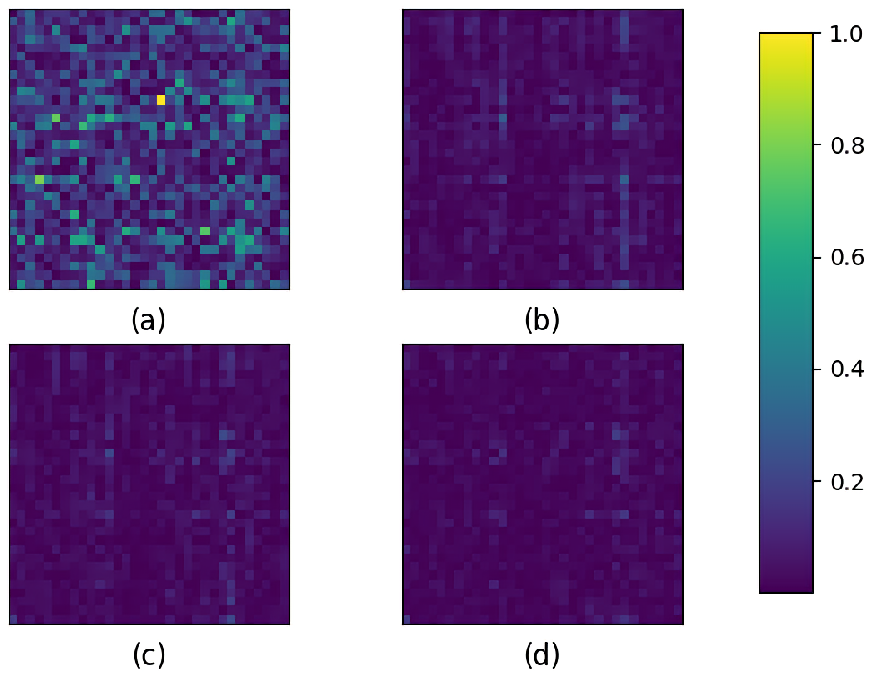}
\DeclareGraphicsExtensions.
\caption{A visualization of a well-trained SC-Attention Network. (a) $\boldsymbol{S_0}$; (b) $\boldsymbol{S}_{2}$; (c) $\boldsymbol{S}_{4}$; (d) $\boldsymbol{S}_{8}$. }
\label{fig:layers}
\vspace{-5pt}
\end{figure}

Due to the existing research work in single-RIS aided systems and the limitation of the space in this letter, we mainly present the results of double reflection link $\boldsymbol{H}_{3k}$ in the following content. In order to understand the effectiveness of the attention block clearly, we visualize the outputs of the SC-attention network with different numbers of attention blocks and the brighter area means more severe noise in Fig. \ref{fig:layers}. As such, $\boldsymbol{S_0}$ is the visualization of the input, while $\boldsymbol{S}_{2}$, $\boldsymbol{S}_{4}$ and $\boldsymbol{S}_{8}$ are the visualizations of the outputs of the SC-attention network, which consist of two, four, and eight attention blocks. The light points in (a) indicate a large amount of noise in CSI matrices. Subsequently, it can be observed that the noise becomes smaller from (b), (c), and (d), which means that the denoising effect is more evident with the increment of the number of the attention blocks. Therefore, Fig. \ref{fig:layers} indicates that the proposed SC-attention method can achieve a satisfactory performance through a block-by-block denoising mechanism. 

\begin{table}[!ht]
\centering
\renewcommand{\tablename}
\caption{\centering{ TABLE I} \protect \\  Accuracy Improvement on Skip Connection}\\
\resizebox{.9\columnwidth}{!}{
\begin{tabular}{ccccccc}
   \toprule
   SNR(dB) & -10 & -5 & 0 & 5 & 10 & 15 \\
   \midrule
   Attention-Only & 0.3842 & 0.2409 & 0.2590 & 0.0939 & 0.0522 & 0.0312 \\
   SC-Attention   & 0.3232 & 0.2200 & 0.1567 & 0.0930  & 0.0508 & 0.0294\\
   \bottomrule
\end{tabular}}
\vspace{-8pt}
\end{table}

 As for the effectiveness of skip-connection structure, Table I compares the NMSE of the double-reflection link $\boldsymbol{H}_{3k}$  in the attention-only network with that in the SC-attention network. Six levels of SNR are investigated in this table, from -10  dB to 15 dB with an interval of 5dB.
 We can observe that skip-connection structure can improve the accuracy of $\boldsymbol{H}_{3k}$ in various SNR environments, especially in the case of SNR $= -10$ dB, the accuracy can be directly improved up to $16\%$.

\section{Conclusion}
We proposed the SC-attention network to improve the estimation performance for both the single-reflection channels and the double-reflection channel in a double-RIS aided massive MIMO network. In the proposed scheme, the recovery of the CSI matrices was modeled as a denoising problem from noisy observations. By applying the skip-connection structure and self-attention layers, the estimation performance was shown to outperform both the conventional and existing DL-based estimators in terms of the NMSE criterion. The denoising effect of the SC-attention network with different numbers of attention blocks is also visualized and the skip-connection structure was shown to improve the accuracy up to $16\%$ with less computational complexity. Moreover, our proposed SC-attention network can be introduced into other problems, such as modulation recognition, signal detection and so on. This is left for our future work.

\vspace{-6pt}

\bibliographystyle{IEEEtran}

\bibliography{IEEEabrv,myref}

\end{document}